\documentclass[reprint,aps, prb, twocolumn, longbibliography, showpacs,preprintnumbers, superscriptaddress, amsmath, amssymb]{revtex4-2}

\usepackage{placeins}
\usepackage{amssymb}
\usepackage{makecell}
\usepackage{mathrsfs}
\usepackage{amsmath}
\usepackage{graphicx}
\usepackage{epstopdf}
\usepackage{float}
\usepackage{multirow}
\usepackage{array}
\usepackage{booktabs}
\usepackage{ulem}
\usepackage{color}
\usepackage{dcolumn}
\usepackage{bm}
\usepackage[colorlinks=true, letterpaper=true, pdfstartview=FitV, linkcolor=blue, citecolor=blue, urlcolor=blue]{hyperref}

\makeatletter
\newcommand{\Rmnum}[1]{\expandafter\@slowromancap\romannumeral #1@}
\makeatother
\usepackage{amsmath}

\begin{document}


\title{Matrix-product entanglement characterizing the optimality of state-preparation quantum circuits}

\author{Shuo Qi}
\affiliation{Department of Physics, Capital Normal University, Beijing 100048, China}

\author{Wen-Jun Li}
\email[Corresponding author. Email: ]{liwenjun18@mails.ucas.ac.cn}
\affiliation{Artificial intelligence College of Putian University, Putian, Fujian, 351100, China}

\author{Gang Su}
\email[Corresponding author. Email: ]{gsu@ucas.ac.cn}
\affiliation{Institute of Theoretical Physics, Chinese Academy of Sciences, Beijing 100190, China}
\affiliation{Kavli Institute for Theoretical Sciences, University of Chinese Academy of Sciences, Beijing 100190, China}

\author{Shi-Ju Ran}
\email[Corresponding author. Email: ]{sjran@cnu.edu.cn}
\affiliation{Department of Physics, Capital Normal University, Beijing 100048, China}

\date{\today}

\begin{abstract}
Multipartite entanglement offers a powerful framework for understanding the complex collective phenomena in quantum many-body systems that are often beyond the description of conventional bipartite entanglement measures. Here, we propose a class of multipartite entanglement measures that incorporate the matrix product state (MPS) representation, enabling the characterization of the optimality of quantum circuits for state preparation. These measures are defined as the minimal distances from a target state to the manifolds of MPSs with specified virtual bond dimensions $\chi$, and thus are dubbed as $\chi$-specified matrix product entanglement ($\chi$-MPE). We demonstrate superlinear, linear, and sublinear scaling behaviors of $\chi$-MPE with respect to the negative logarithmic fidelity $F$ in state preparation, which correspond to excessive, optimal, and insufficient circuit depth $D$ for preparing $\chi$-virtual-dimensional MPSs, respectively. Specifically, a linearly-growing $\chi$-MPE with $F$ suggests $\mathcal{H}_{\chi} \simeq \mathcal{H}_{D}$, where $\mathcal{H}_{\chi}$ denotes the manifold of the $\chi$-virtual-dimensional MPSs and $\mathcal{H}_{D}$ denotes that of the states accessible by the $D$-layer circuits. We provide an exact proof that $\mathcal{H}_{\chi=2} \equiv \mathcal{H}_{D=1}$. Our results establish tensor networks as a powerful and general tool for developing parametrized measures of multipartite entanglement. The matrix product form adopted in $\chi$-MPE can be readily extended to other tensor network ansätze, whose scaling behaviors are expected to assess the optimality of quantum circuit in preparing the corresponding tensor network states.
\end{abstract}

\maketitle

\paragraph*{Introduction.---} Characterizing the optimality of quantum circuits for quantum tasks is a foundational challenge, as it directly influences the efficiency and scalability of quantum algorithms. In quantum state preparation, for instance, minimizing resource consumption, such as gate count and circuit depth~\cite{PhysRevLett.129.230504}, is critical for mitigating noise, enhancing fidelity, and enabling practical implementations on near-term quantum hardware. Entanglement, a cornerstone of quantum advantage, plays a central role in assessing circuit optimality. However, studying entanglement in this context presents significant theoretical and computational challenges, particularly in developing effective entanglement measures.

Bipartite entanglement entropy~\cite{RevModPhys.81.865} is among the most widely used measures, serving as a powerful tool for quantifying quantum computational capabilities~\cite{doi:10.1098/rspa.2002.1097}, detecting quantum phase transitions~\cite{PhysRevA.66.032110,Osterloh2002-ns,PhysRevLett.93.250404}, and more. Yet quantum many-body systems demand descriptions rooted in collective multi-qubit effects, necessitating a shift from bipartite to multipartite entanglement~\cite{10519863}. This shift has been pivotal in advancing quantum protocols such as secret sharing~\cite{PhysRevA.59.1829,PhysRevLett.121.150502} and error correction~\cite{doi:10.1126/science.1131563,Dür_2007,PhysRevLett.134.210602}, while also driving interest in condensed matter physics, where multipartite entanglement characterizes complex interacting systems~\cite{RevModPhys.80.517,DeChiara_2018}.  

Well-established measures of multipartite entanglement include Three-Tangle~\cite{Eltschka_2008,PhysRevLett.102.250404}, global entanglement~\cite{10.10631.1497700,Love2007-pt}, and geometric entanglement (GE)~\cite{PhysRevA.68.042307,PhysRevLett.100.130502,Orús_2014}. GE, defined as the distance between a quantum state and the manifold of product states, offers an intuitive measure of multipartite entanglement. It has found applications in characterizing mixed states~\cite{PhysRevA.68.042307} and quantifying entanglement dimensionality~\cite{PhysRevA.110.012452}. However, GE’s utility is constrained by the inherent limitations of the product state manifold it considers.

\begin{figure}[tbp]
\includegraphics[width=0.96\linewidth]{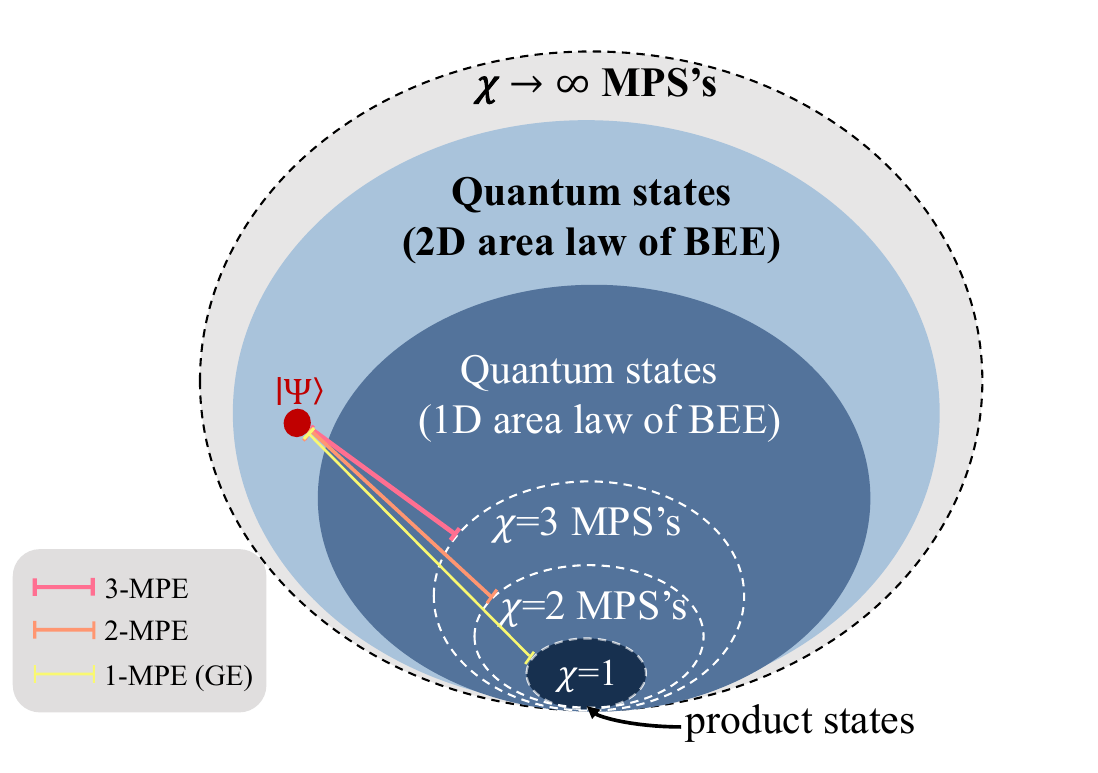}
\caption{\label{fig1} (Color online) Illustration of $\chi$-MPE. The circles indicate different manifolds. The dash circles indicate the manifolds formed by the MPSs with specified virtual bond dimension $\chi$. The solid lines indicate the $\chi$-MPE as the minimal distances from the target state $|\Psi\rangle$ to the corresponding manifolds. For $\chi=1$, the MPSs are product states, and the $1$-MPE reduces to GE.}
\end{figure}

In this work, we employ the matrix product state (MPS) representation~\cite{Schollwoeck2011,O14TNSRev} and propose a class of multipartite entanglement measures. The key idea is measuring multipartite entanglement as the minimal distance to the manifold formed by the MPSs with a specified virtual bond dimension $\chi$. The bipartite entanglement entropy of the quantum states therein obeys the one-dimensional area law~\cite{Hastings_2007,Eisert2010,PhysRevB.78.024410} and is bounded as $S \leq \ln \chi$. We dub such measures as $\chi$-specified matrix product entanglement ($\chi$-MPE), which can be regarded as the generalization of GE by considering the minimal distance to the manifold formed by entangled states with bounded entanglement. Specifically, the $\chi$-MPE is reduced to GE for $\chi=1$. See the illustration in Fig.~\ref{fig1}. 


We further reveal the physical significance of $\chi$-MPE by considering quantum state preparation using variational quantum circuits (VQCs)~\cite{CAB+21VQA}. Specifically, the scaling behavior of $\chi$-MPE is shown to characterize the optimality of state-preparation VQCs' depth. We demonstrate that by increasing $\chi$ in $\chi$-MPE, the super-linear, linear, and sub-linear scaling behaviors of $\chi$-MPE are uncovered against the negative logarithmic fidelity (NLF) $F$~\cite{math10060940,PhysRevLett.100.080601,PhysRevB.103.075106} in state preparation. These three types of behaviors suggest the depth of VQC is excessive, optimal, and insufficient, respectively, for preparing the $\chi$-virtual-dimensional MPSs. A linearly-growing $\chi$-MPE suggests $\mathcal{H}_{\chi} \simeq \mathcal{H}_{D}$, where $\mathcal{H}_{\chi}$ denotes the manifold formed by the $\chi$-virtual-dimensional MPSs and $\mathcal{H}_{D}$ denotes that of the states accessible by the $D$-layer VQCs. Exact proof of $\mathcal{H}_{\chi=2} \equiv \mathcal{H}_{D=1}$ is given. Our findings can be readily generalized by defining novel measures of multipartite entanglement using the other tensor network ansatz~\cite{VMC08MPSPEPSRev, O14TNSRev, RTPC+17TNrev}, whose scaling behaviors are expected to reveal the optimality of VQCs on preparing the corresponding tensor network states.

\paragraph*{Matrix product entanglement.---}
GE (denoted as $E_{\text{GE}}$) is a widely-used measure of multipartite entanglement based on the minimal distance from the target state $|\Psi\rangle$ to the manifold of the product states, satisfying
\begin{equation}
    E_{\text{GE}} =\min_{|\Phi \rangle= \prod_{\otimes i=1}^{N}|\phi^{[i]} \rangle} \left( - \log_{2} | \langle \Psi   | \Phi\rangle | ^{2} \right),
    \label{eq_GE}
\end{equation}
with $|\phi^{[i]} \rangle$ representing a state of the $i$-th qubit. Obviously, GE adopts fidelity distance as the measure and satisfies the triangle inequality~\cite{nielsen2010quantum,PhysRevA.68.042307,PhysRevLett.100.130502,Orús_2014}.

Here, we consider the minimal distance to a manifold of the entangled states satisfying certain restrictions. We focus on a special form of states, known as MPS~\cite{Schollwoeck2011,O14TNSRev}, which can faithfully represent a large class of quantum many-body states~\cite{PhysRevB.78.024410}. We adopt the MPS with open boundary condition that is written as
\begin{equation}
    \left |\Phi\right \rangle =\sum_{a_{1}\cdots a_{N-1}}\sum_{s_{1}\cdots s_{N}}A_{s_{1} a_{1}}^{[1]} A_{s_{2} a_{1} a_{2}}^{[2]}\cdots A_{s_{N} a_{N-1}}^{[N]} \prod_{\otimes n=1}^{N} \left | s_{n}  \right \rangle,
    \label{eq_MPS}
\end{equation}
where $\{s_{n}\}$ and $\{a_{n}\}$ are called physical and virtual indexes, respectively, and $N$ denotes the total number of qubits. Note we have $\dim(s_{n})=2$ for qubits (namely spin-$1/2$ particles). 

For $\{a_{n}\}$, we typically impose an upper bound on their dimensions and call it virtual dimension (denoted as $\chi$). Such a dimension determines the parameter complexity of the MPS and the maximal bipartite entanglement entropy that the MPS can carry. In this sense, we can naturally introduce a special manifold using the MPSs with a fixed $\chi$, and define the $\chi$-MPE as
\begin{equation}
\begin{split}
    E_{\chi} =\min_{|\Phi_{\chi}\rangle} \left(-\log_2 \left | \langle \Psi   |\Phi_{\chi}\rangle  \right | ^{2} \right),
    \label{eq_TTGE}
\end{split}
\end{equation}
where $|\Phi_{\chi}\rangle$ denotes an MPS with $\dim(a_n) \leq \chi$ for any $n$. This minimization problem is known as tensor-train rounding~\cite{doi:10.1137/090752286,SM}.

MPS possesses some properties that are critical to MPE in the sense of defining manifolds. For $\chi=1$, an MPS reduces to a product state, making $E_{\chi=1}$ equivalent to $E_{\text{GE}}$. For $\chi>1$, $\chi$-MPE measures the minimal distance from the target state to the manifold $\mathcal{H}_{\chi}$ formed by the $\chi$-virtual-dimensional MPSs. We have $\mathcal{H}_{\chi} \subset \mathcal{H}_{\chi'}$ for $1 \leq \chi < \chi'$. As $\chi \to \infty$, the MPS can approach any state, meaning $\mathcal{H}_{\chi \to \infty}$ approaches the entire Hilbert space. Such relations are illustrated in Fig.~\ref{fig1}.

The manifold $\mathcal{H}_{\chi}$ can be characterized using the bipartite entanglement entropy $S$. To this end, we partition the system into two subsystems $\mathcal{A}=(1, \cdots, n')$ and $\mathcal{B}=(n'+1, \cdots, N)$. This leads to the Schmidt number $\mathcal{N} \leq \dim(a_{n'})$, implying the bipartite entanglement entropy bound $S \leq \ln \chi$. This is known as the one-dimensional area law of entanglement entropy~\cite{Hastings_2007,Eisert2010,PhysRevB.78.024410}.

\begin{figure}[tbp]
\includegraphics[width=1.02\linewidth]{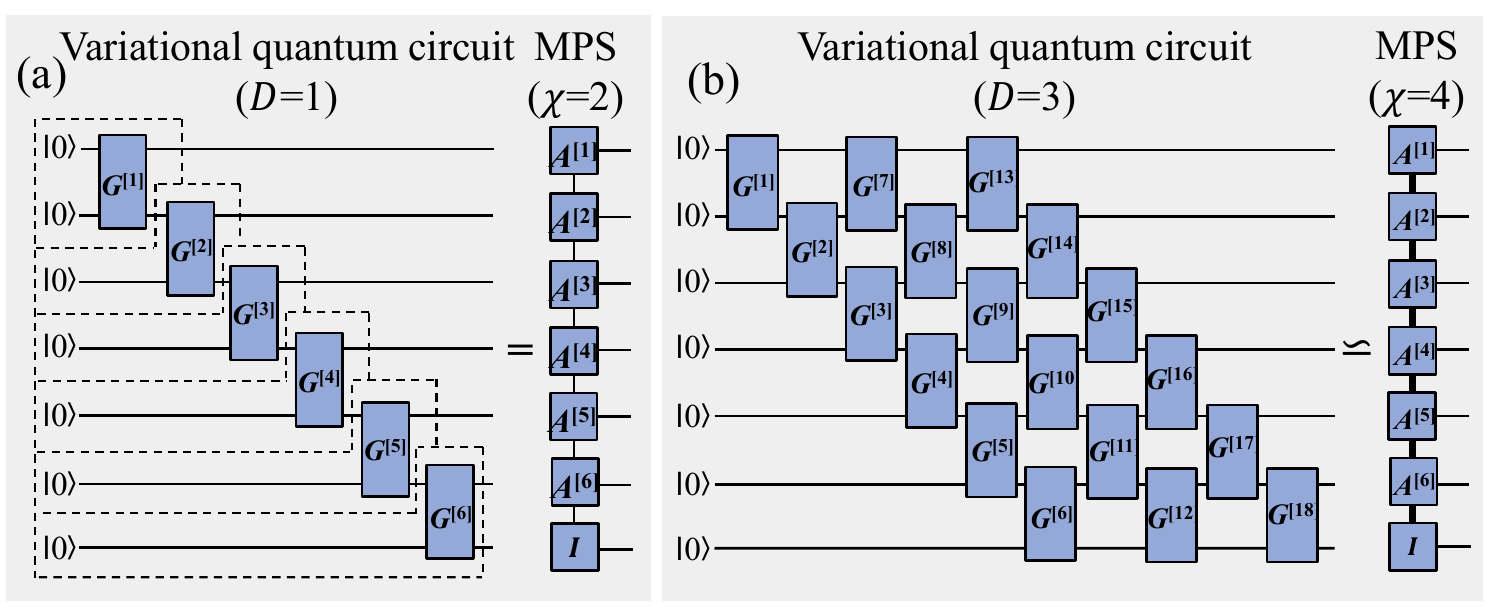}
\caption{\label{fig2}(Color online) (a) The illustration of the equivalence between the MPSs with $\chi=2$ and the quantum states prepared by single-layer VQCs from the product state $|0\cdots0\rangle$. (b) Our numerical results suggest the MPSs with $\chi=4$ can be efficiently prepared using a VQC with 3 layers.}
\end{figure}

\paragraph*{Exact results for $\chi=2$.---}
Below, we prove the equivalence between $\mathcal{H}_{\chi=2}$ and the manifold of the states accessible using single-layer VQC from product state (denoted as $\mathcal{H}_{D=1}$), namely $\mathcal{H}_{\chi=2} \equiv \mathcal{H}_{D=1}$. For simplicity, we consider the stair-like circuit as shown in Fig.~\ref{fig2}(a).

Denoting the quantum gates in the circuit as $\{\hat{G}^{[n]}\}$ ($n=1, \cdots, N-1$), we define the tensors as
\begin{eqnarray}
    &&A^{[1]}_{s_1 a_1} = \langle 00 |\hat{G}^{[1]}|s_1 a_1\rangle, \label{eq-A1} \\
    &&A^{[n]}_{s_n a_{n
    -1} a_n}  = \langle a_{n-1} 0 |\hat{G}^{[n]}|s_n a_n\rangle \ (2 \leq n \leq N-1), \label{eq-An} \\
    &&A^{[N]}_{s_N a_{N-1}} = I_{s_N a_{N-1}} \label{eq-AN},
    \label{eq_A}
\end{eqnarray}
with $I$ denoting an identity and $|\ast\rangle$ a two-level state of a qubit. In this construction, the resulting tensors form a $\chi=2$ MPS [see Eq.~(\ref{eq_MPS})]. Notably, some degrees of freedom of the qubits during the evolution are treated as the virtual degrees of freedom of the MPS. Therefore, these virtual bond dimensions equal to the dimension of a qubit, i.e., $\chi=\dim(s_{n})=2$. Eqs.~(\ref{eq-A1})-(\ref{eq-AN}) mean that the local tensors $\{A^{[n]}\}$ forming a $\chi=2$ MPS can be completely determined by the quantum gates of the single-layer circuit. Therefore, the state prepared by any given single-layer circuit can be written as an MPS, i.e., $\mathcal{H}_{D=1} \subseteq \mathcal{H}_{\chi=2}$.

Below we show how to determine the corresponding quantum gates, provided by the local tensors $\{A^{[n]}\}$ of a $\chi=2$ MPS. Part of the elements in the quantum gates can be directly determined according to Eqs.~(\ref{eq-A1})-(\ref{eq-AN}). The rest elements of the gates can be chosen as the vectors in the kernel space based on the orthogonality of the quantum gates~\cite{Ran2020a}. 

Considering to determine $\hat{G}^{[n]}$ ($2 \leq n \leq N-1$) from $A^{[n]}_{s_1 a_{n-1} a_n}$ as an example, the elements $\langle a_{n-1} a |\hat{G}^{[n]}|s_n a_n\rangle$ for $a=0$ are specified by Eqs.~(\ref{eq-A1})-(\ref{eq-AN}). The remaining task is to obtain the elements $\langle a_{n-1} a |\hat{G}^{[n]}|s_n a_n\rangle$ for $a=1$. In practical simulations, we can define the matrix $M_{s'_n a'_n, s_n a_n} = \sum_{a_{n-1}} \langle s'_n a'_n|\hat{G}^{[n]\dagger}|a_{n-1} 0 \rangle \langle a_{n-1} 0 |\hat{G}^{[n]}|s_n a_n\rangle$, which is rank-deficient, namely $\text{rank}(M) = 2 < 4$. Within the subspace spanned by the two eigenvectors with zero eigenvalues of $M$, any two mutually-orthogonal vector, denoted as $v_{s_na_n}$ and $v'_{s_na_n}$, provide a solution. Specifically, we may have $\langle 0 1 |\hat{G}^{[n]}|s_n a_n\rangle = v_{s_na_n}$ and $\langle 1 1 |\hat{G}^{[n]}|s_n a_n\rangle = v'_{s_na_n}$. Such a construction guarantees the unitarity of the gates since $\hat{G}^{[n]}$ is formed by normalized vectors that are mutually orthogonal. Therefore, one can always obtain the quantum gates of a single-layer circuit to prepare an arbitrarily given MPS with $\chi=2$, and thus $\mathcal{H}_{\chi=2} \subseteq \mathcal{H}_{D=1}$. 

Summarizing the above results, we have $\mathcal{H}_{\chi=2} \equiv \mathcal{H}_{D=1}$~\cite{SM}. One physical consequence is that the 2-MPE, which measures the minimal distance to the manifold $\mathcal{H}_{\chi=2}$, also gives the minimal distance to $\mathcal{H}_{D=1}$. Below, we utilize such a relation and propose to use $\chi$-MPE for measuring VQC's optimality on preparing MPSs.

\begin{figure*}[tbp]
\centering
\includegraphics[width=17.8cm]{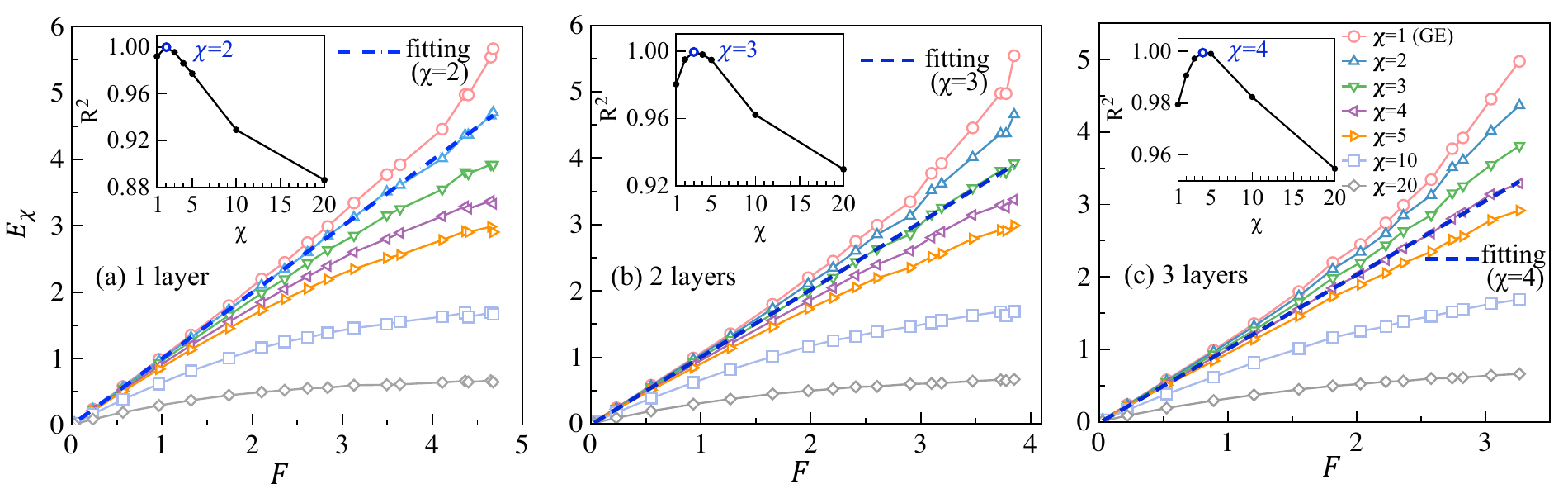}
\caption{\label{fig3}(Color online) The $\chi$-MPE $E_{\chi}$ [Eq.~(\ref{eq_TTGE})] versus the NLF $F$ [Eq.~(\ref{eq_NLF})] for (a) $D=1$, (b) $D=2$, and (c) $D=3$, with $D$ the number of layers of the VQC. We take the $N$-qubit generalized RPSs as the target states with $N=12$. The insets show the coefficient of determination ($R^2$) that characterizes the linearity of the relation between $E_{\chi}$ and $F$ for different values of $\chi$.}
\FloatBarrier
\end{figure*}

\paragraph*{Numerical results with $D=1$.---} The state-preparation accuracy can be characterized by the NLF, which is defined as
\begin{equation}
    F=-\log_2 \left | \left \langle \Psi | \Phi_{D}   \right \rangle  \right |^2,
    \label{eq_NLF}
\end{equation}
with $|\Phi_{D}\rangle$ the state prepared by a $D$-layer VQC~\cite{SM}. The equivalence between $\mathcal{H}_{\chi=2}$ and $\mathcal{H}_{D=1}$ should render a linear relationship between the 2-MPE and NLF with $D=1$.


Below, we take $|\Psi\rangle$ as generalized random pure states (RPSs)~\cite{Weedbrook2012} and show the $\chi$-MPE $E_{\chi}$ [Eq.~(\ref{eq_TTGE})] versus the NLF $F$ [Eq.~(\ref{eq_NLF})] with a single-layer stair-like VQC ($D=1$). We adopt the automatically-differentiable tensors to parameterize and optimize the VQC~\cite{PhysRevA.104.042601}. The elements of a generalized RPS are generated by sampling according to the Gaussian distribution $\mathcal{N}(\mu, \sigma)$ with $\mu$ and $\sigma$ representing the mean and standard deviation, respectively. Taking $\mu=0$ and $\sigma=1$, one obtains the standard RPS whose bipartite entanglement entropy obeys the volume law as $S = \log_2 2^{N/2} - \frac{1}{2\ln2}$~\cite{PhysRevLett.71.1291,PRXQuantum.3.030201}. Therefore, RPS satisfies the volume law of entanglement entropy and should give non-vanishing $E_2$ and $F$. In other words, RPSs are generally outside the manifolds $\mathcal{H}_{D}$ and $\mathcal{H}_{\chi}$.

We here obtain the generalized RPSs with different entanglement strength by varying $\sigma$ while fixing $\mu=5$. These states are used to demonstrate the scaling relation between $\chi$-MPE and NLF. The results for $D=1$ show a linear relation between $E_{2}$ and $F$ as
\begin{equation}
    E_{2} \simeq kF,
    \label{eq_linear}
\end{equation}
with $k \simeq 1$. This is consistent with our conclusion on $\mathcal{H}_{\chi=2} \equiv \mathcal{H}_{D=1}$. 

If we reduce $\chi$ to 1, $E_{\chi}$ exhibits a super-linear relationship to $F$, as indicated by the red circles in Fig.~\ref{fig3}(a). This means the minimal distance from the target state to $\mathcal{H}_{D=1}$ is smaller than that to $\mathcal{H}_{\chi=1}$. Therefore, we suggest the super-linear relation as a signature on the excessive depth of the VQC for preparing the $\chi$-virtual dimensional MPSs. 

For $\chi>2$, our results show the sub-linear growth of $E_{\chi}$ versus $F$ with a single-layer VQC. This can serve as a signature on the insufficiency of the VQC's depth for preparing the MPSs with the corresponding virtual dimension $\chi$. The linearity of the relationship between $F$ and $E_{\chi}$, which is characterized by the coefficient of determination $R^2$, reaches its maximum at $\chi=2$ [see the inset of Fig.~\ref{fig3}(a) with $R^2 > 0.999$]. The maximal linearity suggests that VQC possesses an optimal depth for preparing the MPSs with the corresponding virtual dimension. 

We shall stress that the generalized RPSs here play the role as reference states. The $\chi$-MPE, which though measures the multipartite entanglement of the generalized RPS, is here mainly used to reveal the underlying relation between the MPS and VQC for state preparation. Other states outside the manifolds ($\mathcal{H}_{D}$ and $\mathcal{H}_{\chi}$) should also lead to the same scaling behavior of $E_{\chi}$ against $F$, which can be similarly used to characterize the optimality of VQC's depth on preparing MPSs.

\paragraph*{Numerical results with $D>1$.---} Deeper VQCs should optimally prepare the MPSs with larger $\chi$, as illustrated in Fig.~\ref{fig2}(b). Fig.~\ref{fig3}(b) and (c) show the the scaling behavior of $E_{\chi}$ against $F$ for $D=2$ and $3$, respectively. As $\chi$ increases, the growth of $E_{\chi}$ alters asymptotically from being super-linear, linear, and finally to sub-linear. The peak of $R^2$ shown in the insets indicates the ``critical'' $\chi$ that renders a linear relation between $E_{\chi}$ and $F$.


\begin{figure}[tbp]
\includegraphics[width=0.95\linewidth]{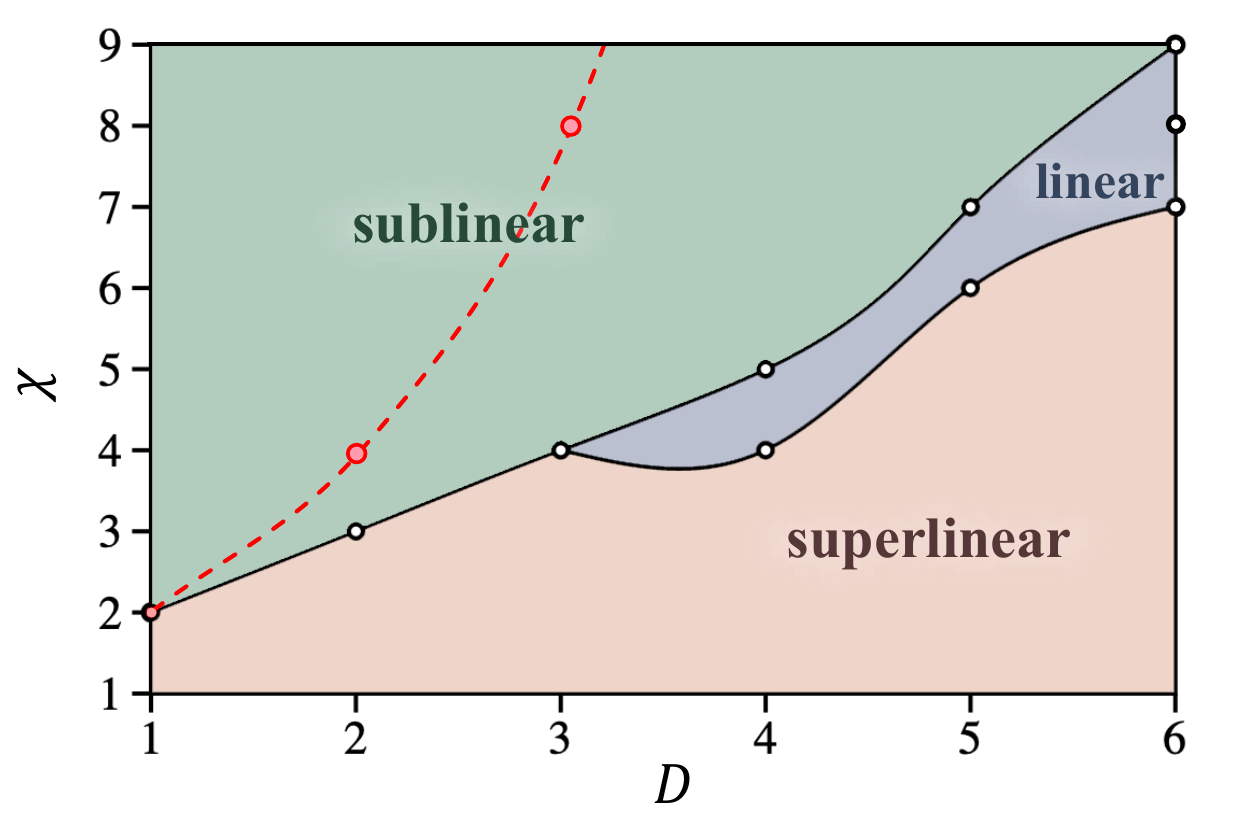}
\caption{\label{fig4} (Color online) The scaling diagram demonstrating three regions, where the $\chi$-MPE exhibits superlinear, linear, and sublinear behaviors versus the NLF with $D$-layer VQCs, respectively. Note we consider the scaling to be linear with the coefficient of determination $R^2> 0.999$. The black lines are drawn as curves using cubic spline interpolation between integer-valued data points, with the intention solely for an artistic illustration. The red dashed line shows the theoretical upper bounds $\chi=2^{D}$~\cite{SM}.}
\end{figure}

Figure~\ref{fig4} shows the ``scaling" diagram of $\chi$-MPE for different $\chi$ and depth $D$ of the VQCs. Here, we deem the scaling relation to be linear for $R^2 > 0.999$. For $D >3$, we obtain multiple values of $\chi$ that give $R^2 > 0.999$. These results reveal the ``non-equivalent non-subordinate'' relationship between two kinds of manifolds defined by, respectively, the MPSs with a specified virtual dimension and the states prepared by the VQCs with a specified depth. Note that the upper boundary of the linear region in Fig.~\ref{fig4} is much lower than the theoretical upper bounds $\chi=2^{D}$ obtained by the exponential increase of the virtual dimension in the implementation of VQC to an MPS~\cite{SM}. Considering $\mathcal{H}_{\chi} \subset \mathcal{H}_{\chi'}$ for $\chi < \chi'$, a $D$-layer VQC can accurately prepare a $\chi$-virtual-dimensional MPS if the scaling behavior of $E_{\chi'}$ is linear. Thus, one may take the maximal $D$ as a sufficient depth of the VQC to prepare an MPS with a specified $\chi$.

\paragraph*{Summary and perspective.---}
In this work, we propose a novel class of multipartite entanglement measures termed $\chi$-specified matrix product entanglement ($\chi$-MPE), which quantifies the minimal distance from a target state to the manifold spanned by matrix product states (MPSs) with virtual bond dimension $\chi$. For $\chi=1$, MPSs reduce to product states, rendering $\chi$-MPE equivalent to standard geometric entanglement. For $\chi=2$, we rigorously prove the equivalence between the manifold of $\chi=2$ MPSs and that generated by single-layer variational quantum circuits (VQCs). This equivalence establishes a linear relationship between 2-MPE and the negative logarithmic fidelity (NLF) in state preparation, serving as a hallmark of optimal VQC depth.  

Extending our analysis to multi-layer VQCs, we uncover asymptotic scaling transitions in $\chi$-MPE versus NLF. As $\chi$ increases, the scaling behavior transitions from super-linear to linear and finally to sub-linear, corresponding to excessive, optimal, and insufficient VQC depths for MPS preparation, respectively. These results establish $\chi$-MPE as a critical metric for VQC optimization, bridging entanglement theory, tensor networks, and quantum computational complexity. The framework readily generalizes to other tensor network ansätze, enabling efficiency assessments of VQCs in preparing diverse tensor network states.

\textit{Acknowledgments.} This work was supported in part by the Innovation Program for Quantum Science and Technology (Grant No. 2024ZD0300500), NSFC (Grant No. 12534009 and 12447101), Beijing Natural Science Foundation (Grant No. 1232025), the Strategic Priority Research Program of Chinese Academy of Sciences (Grant No.XDB1270000) and CAS. S.J.R. was supported in part by the Ministry of Education Key Laboratory of Quantum Physics and Photonic Quantum Information (Grant No. ZYGX2024K020), the Peng Huanwu Visiting Professor Program, CAS, and Academy for Multidisciplinary Studies, Capital Normal University. W.J.L. is supported by the Introduction of Talents to Start Scientific Research Projects in Putian University under Grant 2024143, and Putian Science and Technology Plan Project under Grant 2023GJGZ003. The numerical simulations were partially performed on the robotic AI-Scientist platform of Chinese Academy of Sciences.

\normalem
\bibliography{MPE.bib}

@article{PhysRevA.59.1829,
  title = {Quantum secret sharing},
  author = {Hillery, Mark and Bifmmode Buzek, Vladimr and Berthiaume, Andre},
  journal = {Phys. Rev. A},
  volume = {59},
  issue = {3},
  pages = {1829--1834},
  numpages = {0},
  year = {1999},
  month = {Mar},
  publisher = {American Physical Society},
  doi = {10.1103/PhysRevA.59.1829},
}

@article{PhysRevA.68.042307,
  title = {Geometric measure of entanglement and applications to bipartite and multipartite quantum states},
  author = {Wei, Tzu-Chieh and Goldbart, Paul M.},
  journal = {Phys. Rev. A},
  volume = {68},
  issue = {4},
  pages = {042307},
  numpages = {12},
  year = {2003},
  month = {Oct},
  publisher = {American Physical Society},
  doi = {10.1103/PhysRevA.68.042307},
  url = {https://link.aps.org/doi/10.1103/PhysRevA.68.042307}
}

@article{RevModPhys.81.865,
  title = {Quantum entanglement},
  author = {Horodecki, Ryszard and Horodecki, Pawel and Horodecki, Michal and Horodecki, Karol},
  journal = {Rev. Mod. Phys.},
  volume = {81},
  issue = {2},
  pages = {865--942},
  numpages = {0},
  year = {2009},
  month = {Jun},
  publisher = {American Physical Society},
  doi = {10.1103/RevModPhys.81.865},
  url = {https://link.aps.org/doi/10.1103/RevModPhys.81.865}
}

@INBOOK{10519863,
  author={Walter, Michael and Gross, David and Eisert, Jens},
  booktitle={Quantum Information: From Foundations to Quantum Technology Applications}, 
  title={Multipartite Entanglement}, 
  year={2019},
  volume={},
  number={},
  pages={293-330},
  doi={10.1002/9783527805785.ch14}}

@article{PhysRevA.104.042601,
  title = {Automatically differentiable quantum circuit for many-qubit state preparation},
  author = {Zhou, Peng-Fei and Hong, Rui and Ran, Shi-Ju},
  journal = {Phys. Rev. A},
  volume = {104},
  issue = {4},
  pages = {042601},
  numpages = {7},
  year = {2021},
  month = {Oct},
  publisher = {American Physical Society},
  doi = {10.1103/PhysRevA.104.042601},
  url = {https://link.aps.org/doi/10.1103/PhysRevA.104.042601}
}

@article{PhysRevLett.93.250404,
  title = {Quantum Phase Transitions and Bipartite Entanglement},
  author = {Wu, L.-A. and Sarandy, M. S. and Lidar, D. A.},
  journal = {Phys. Rev. Lett.},
  volume = {93},
  issue = {25},
  pages = {250404},
  numpages = {4},
  year = {2004},
  month = {Dec},
  publisher = {American Physical Society},
  doi = {10.1103/PhysRevLett.93.250404},
  url = {https://link.aps.org/doi/10.1103/PhysRevLett.93.250404}
}

@Article{Eisert2010,
  author    = {Eisert, J and Cramer, M and Plenio, M B},
  journal   = {Rev. Mod. Phys.},
  title     = {Colloquium: Area laws for the entanglement entropy},
  year      = {2010},
  month     = feb,
  number    = {1},
  pages     = {277--306},
  volume    = {82},
  copyright = {http://link.aps.org/licenses/aps-default-license},
  language  = {en},
  publisher = {American Physical Society (APS)},
  doi = { https://doi.org/10.1103/RevModPhys.82.277},
}

@Article{Weedbrook2012,
  author    = {Weedbrook, Christian and Pirandola, Stefano and Garcia-Patron, Raul and Cerf, Nicolas J and Ralph, Timothy C and Shapiro, Jeffrey H and Lloyd, Seth},
  journal   = {Rev. Mod. Phys.},
  title     = {Gaussian quantum information},
  year      = {2012},
  month     = may,
  number    = {2},
  pages     = {621--669},
  volume    = {84},
  copyright = {http://link.aps.org/licenses/aps-default-license},
  language  = {en},
  publisher = {American Physical Society (APS)},
  doi = { https://doi.org/10.1103/RevModPhys.84.621},
}

@Article{Ran2020a,
  title = {Encoding of matrix product states into quantum circuits of one- and two-qubit gates},
  author = {Ran, Shi-Ju},
  journal = {Phys. Rev. A},
  volume = {101},
  issue = {3},
  pages = {032310},
  numpages = {7},
  year = {2020},
  month = {Mar},
  publisher = {American Physical Society},
  doi = {10.1103/PhysRevA.101.032310},
  url = {https://link.aps.org/doi/10.1103/PhysRevA.101.032310}
}

@article{RevModPhys.80.517,
  title = {Entanglement in many-body systems},
  author = {Amico, Luigi and Fazio, Rosario and Osterloh, Andreas and Vedral, Vlatko},
  journal = {Rev. Mod. Phys.},
  volume = {80},
  issue = {2},
  pages = {517--576},
  numpages = {0},
  year = {2008},
  month = {May},
  publisher = {American Physical Society},
  doi = {10.1103/RevModPhys.80.517},
  url = {https://link.aps.org/doi/10.1103/RevModPhys.80.517}
}

@Article{Schollwoeck2011,
  author    = {Schollwock, Ulrich},
  journal   = {Ann. Phys. (N. Y.)},
  title     = {The density-matrix renormalization group in the age of matrix product states},
  year      = {2011},
  month     = jan,
  number    = {1},
  pages     = {96--192},
  volume    = {326},
  language  = {en},
  publisher = {Elsevier BV},
  doi = {https://doi.org/10.1016/j.aop.2010.09.012},
}

@article{10.10631.1497700,
    author = {Meyer, David A. and Wallach, Nolan R.},
    title = {Global entanglement in multiparticle systems},
    journal = {Journal of Mathematical Physics},
    volume = {43},
    number = {9},
    pages = {4273-4278},
    year = {2002},
    month = {09},
    doi = {https://doi.org/10.1063/1.1497700},
}

@ARTICLE{Osterloh2002-ns,
  title     = "Scaling of entanglement close to a quantum phase transition",
  author    = "Osterloh, A and Amico, Luigi and Falci, G and Fazio, Rosario",
  journal   = "Nature",
  publisher = "Springer Science and Business Media LLC",
  volume    =  416,
  number    =  6881,
  pages     = "608--610",
  month     =  apr,
  year      =  2002,
  language  = "en",
  doi = { https://doi.org/10.1103/PhysRevB.89.134101},
}

@article{DeChiara_2018,
doi = {10.1088/1361-6633/aabf61},
url = {https://dx.doi.org/10.1088/1361-6633/aabf61},
year = {2018},
month = {jun},
publisher = {IOP Publishing},
volume = {81},
number = {7},
pages = {074002},
author = {De Chiara, Gabriele and Sanpera, Anna},
title = {Genuine quantum correlations in quantum many-body systems: a review of recent progress},
journal = {Reports on Progress in Physics},
}

@Article{CAB+21VQA,
author={Cerezo, M.
and Arrasmith, Andrew
and Babbush, Ryan
and Benjamin, Simon C.
and Endo, Suguru
and Fujii, Keisuke
and McClean, Jarrod R.
and Mitarai, Kosuke
and Yuan, Xiao
and Cincio, Lukasz
and Coles, Patrick J.},
title={Variational quantum algorithms},
journal={Nature Reviews Physics},
year={2021},
month={Sep},
day={01},
volume={3},
number={9},
pages={625-644},
issn={2522-5820},
doi={10.1038/s42254-021-00348-9},
url={https://doi.org/10.1038/s42254-021-00348-9}
}

@book{RTPC+17TNrev,
    author = {Ran, Shi-Ju and Tirrito, Emanuele and Peng, Cheng and Chen, Xi and Tagliacozzo, Luca and Su, Gang and Lewenstein, Maciej},
    publisher = {Springer, Cham},
    doi = {10.1007/978-3-030-34489-4},
    title = {Tensor Network Contractions: Methods and Applications to Quantum Many-Body Systems},
    year = {2020}
}

@article{VMC08MPSPEPSRev,
    author = {Verstraete, Frank and Murg, Valentin and Cirac, J. Ignacio},
    journal = {Advances in Physics},
    pages = {143--224},
    doi = {10.1080/14789940801912366},
    title = {{Matrix product states, projected entangled pair states, and variational renormalization group methods for quantum spin systems}},
    volume = {57},
    year = {2008}
}

@article{O14TNSRev,
    author = {Orus, Roman},
    journal = {Ann. Phys.},
    pages = {117},
    title = {{A practical introduction to tensor networks: Matrix product states and projected entangled pair states}},
    volume = {349},
    year = {2014},
    doi = {10.1016/j.aop.2014.06.013}
}

@article{doi:10.1098/rspa.2002.1097,
author = {Jozsa, Richard  and Linden, Noah },
title = {On the role of entanglement in quantum-computational speed-up},
journal = {Proceedings of the Royal Society of London. Series A: Mathematical, Physical and Engineering Sciences},
volume = {459},
number = {2036},
pages = {2011-2032},
year = {2003},
doi = {10.1098/rspa.2002.1097},
}

@article{PhysRevA.66.032110,
  title = {Entanglement in a simple quantum phase transition},
  author = {Osborne, Tobias J. and Nielsen, Michael A.},
  journal = {Phys. Rev. A},
  volume = {66},
  issue = {3},
  pages = {032110},
  numpages = {14},
  year = {2002},
  month = {Sep},
  publisher = {American Physical Society},
  doi = {10.1103/PhysRevA.66.032110},
  url = {https://link.aps.org/doi/10.1103/PhysRevA.66.032110}
}

@article{Dür_2007,
doi = {10.1088/0034-4885/70/8/R03},
year = {2007},
month = {jul},
publisher = {},
volume = {70},
number = {8},
pages = {1381},
author = {Dur, W and Briegel, H J},
title = {Entanglement purification and quantum error correction},
journal = {Reports on Progress in Physics},
}

@article{PhysRevLett.100.130502,
  title = {Universal Geometric Entanglement Close to Quantum Phase Transitions},
  author = {Orus, Roman},
  journal = {Phys. Rev. Lett.},
  volume = {100},
  issue = {13},
  pages = {130502},
  numpages = {4},
  year = {2008},
  month = {Apr},
  publisher = {American Physical Society},
  doi = {10.1103/PhysRevLett.100.130502},
}

@article{Hastings_2007,
doi = {10.1088/1742-5468/2007/08/P08024},
year = {2007},
month = {aug},
publisher = {},
volume = {2007},
number = {08},
pages = {P08024},
author = {Hastings, M B},
title = {An area law for one-dimensional quantum systems},
journal = {Journal of Statistical Mechanics: Theory and Experiment},
}

@book{nielsen2010quantum,
  title={Quantum computation and quantum information},
  author={Nielsen, Michael A and Chuang, Isaac L},
  year={2010},
  publisher={Cambridge university press}
}

@article{PhysRevB.78.024410,
  title = {Scaling of entanglement support for matrix product states},
  author = {Tagliacozzo, L. and de Oliveira, Thiago. R. and Iblisdir, S. and Latorre, J. I.},
  journal = {Phys. Rev. B},
  volume = {78},
  issue = {2},
  pages = {024410},
  numpages = {14},
  year = {2008},
  month = {Jul},
  publisher = {American Physical Society},
  doi = {10.1103/PhysRevB.78.024410},
  url = {https://link.aps.org/doi/10.1103/PhysRevB.78.024410}
}

@article{PhysRevLett.71.1291,
  title = {Average entropy of a subsystem},
  author = {Page, Don N.},
  journal = {Phys. Rev. Lett.},
  volume = {71},
  issue = {9},
  pages = {1291--1294},
  numpages = {0},
  year = {1993},
  month = {Aug},
  publisher = {American Physical Society},
  doi = {10.1103/PhysRevLett.71.1291},
}

@Article{math10060940,
AUTHOR = {Li, Wei-Ming and Ran, Shi-Ju},
TITLE = {Non-Parametric Semi-Supervised Learning in Many-Body Hilbert Space with Rescaled Logarithmic Fidelity},
JOURNAL = {Mathematics},
VOLUME = {10},
YEAR = {2022},
NUMBER = {6},
ARTICLE-NUMBER = {940},
ISSN = {2227-7390},
DOI = {10.3390/math10060940}
}

@article{PhysRevLett.100.080601,
  title = {Ground State Fidelity from Tensor Network Representations},
  author = {Zhou, Huan-Qiang and Orus, Roman and Vidal, Guifre},
  journal = {Phys. Rev. Lett.},
  volume = {100},
  issue = {8},
  pages = {080601},
  numpages = {4},
  year = {2008},
  month = {Feb},
  publisher = {American Physical Society},
  doi = {10.1103/PhysRevLett.100.080601},
}

@article{PhysRevB.103.075106,
  title = {Visualizing quantum phases and identifying quantum phase transitions by nonlinear dimensional reduction},
  author = {Yang, Yuan and Sun, Zheng-Zhi and Ran, Shi-Ju and Su, Gang},
  journal = {Phys. Rev. B},
  volume = {103},
  issue = {7},
  pages = {075106},
  numpages = {11},
  year = {2021},
  month = {Feb},
  publisher = {American Physical Society},
  doi = {10.1103/PhysRevB.103.075106},
}

@article{PhysRevA.110.012452,
  title = {Quantifying subspace entanglement with geometric measures},
  author = {Zhu, Xuanran and Zhang, Chao and Zeng, Bei},
  journal = {Phys. Rev. A},
  volume = {110},
  issue = {1},
  pages = {012452},
  numpages = {11},
  year = {2024},
  month = {Jul},
  publisher = {American Physical Society},
  doi = {10.1103/PhysRevA.110.012452},
}

@article{PRXQuantum.3.030201,
  title = {Volume-Law Entanglement Entropy of Typical Pure Quantum States},
  author = {Bianchi, Eugenio and Hackl, Lucas and Kieburg, Mario and Rigol, Marcos and Vidmar, Lev},
  journal = {PRX Quantum},
  volume = {3},
  issue = {3},
  pages = {030201},
  numpages = {77},
  year = {2022},
  month = {Jul},
  publisher = {American Physical Society},
  doi = {10.1103/PRXQuantum.3.030201},
}

@article{PhysRevLett.129.230504,
  title = {Quantum State Preparation with Optimal Circuit Depth: Implementations and Applications},
  author = {Zhang, Xiao-Ming and Li, Tongyang and Yuan, Xiao},
  journal = {Phys. Rev. Lett.},
  volume = {129},
  issue = {23},
  pages = {230504},
  numpages = {6},
  year = {2022},
  month = {Nov},
  publisher = {American Physical Society},
  doi = {10.1103/PhysRevLett.129.230504},
  url = {https://link.aps.org/doi/10.1103/PhysRevLett.129.230504}
}

@article{PhysRevLett.121.150502,
  title = {Quantum Secret Sharing Among Four Players Using Multipartite Bound Entanglement of an Optical Field},
  author = {Zhou, Yaoyao and Yu, Juan and Yan, Zhihui and Jia, Xiaojun and Zhang, Jing and Xie, Changde and Peng, Kunchi},
  journal = {Phys. Rev. Lett.},
  volume = {121},
  issue = {15},
  pages = {150502},
  numpages = {6},
  year = {2018},
  month = {Oct},
  publisher = {American Physical Society},
  doi = {10.1103/PhysRevLett.121.150502},
}

@article{
doi:10.1126/science.1131563,
author = {Todd Brun  and Igor Devetak  and Min-Hsiu Hsieh },
title = {Correcting Quantum Errors with Entanglement},
journal = {Science},
volume = {314},
number = {5798},
pages = {436-439},
year = {2006},
doi = {10.1126/science.1131563},
}

@article{PhysRevLett.134.210602,
  title = {How Much Entanglement Is Needed for Quantum Error Correction?},
  author = {Bravyi, Sergey and Lee, Dongjin and Li, Zhi and Yoshida, Beni},
  journal = {Phys. Rev. Lett.},
  volume = {134},
  issue = {21},
  pages = {210602},
  numpages = {6},
  year = {2025},
  month = {May},
  publisher = {American Physical Society},
  doi = {10.1103/PhysRevLett.134.210602}
}

@article{Eltschka_2008,
doi = {10.1088/1367-2630/10/4/043014},
url = {https://dx.doi.org/10.1088/1367-2630/10/4/043014},
year = {2008},
month = {apr},
publisher = {},
volume = {10},
number = {4},
pages = {043014},
author = {Eltschka, C and Osterloh, A and Siewert, J and Uhlmann, A},
title = {Three-tangle for mixtures of generalized GHZ and generalized W states},
journal = {New Journal of Physics}
}

@article{PhysRevLett.102.250404,
  title = {Tripartite Entanglement versus Tripartite Nonlocality in Three-Qubit Greenberger-Horne-Zeilinger-Class States},
  author = {Ghose, S. and Sinclair, N. and Debnath, S. and Rungta, P. and Stock, R.},
  journal = {Phys. Rev. Lett.},
  volume = {102},
  issue = {25},
  pages = {250404},
  numpages = {4},
  year = {2009},
  month = {Jun},
  publisher = {American Physical Society},
  doi = {10.1103/PhysRevLett.102.250404}
}

@ARTICLE{Love2007-pt,
  title     = "A characterization of global entanglement",
  author    = "Love, Peter J and van den Brink, Alec Maassen and Smirnov, A Yu
               and Amin, M H S and Grajcar, M and Il'ichev, E and Izmalkov, A
               and Zagoskin, A M",
  journal   = "Quantum Inf. Process.",
  publisher = "Springer Science and Business Media LLC",
  volume    =  6,
  number    =  3,
  pages     = "187--195",
  month     =  jun,
  year      =  2007,
  language  = "en",
doi = {10.1007/s11128-007-0052-7}
}

@article{Orús_2014,
doi = {10.1088/1367-2630/16/1/013015},
url = {https://dx.doi.org/10.1088/1367-2630/16/1/013015},
year = {2014},
month = {jan},
publisher = {IOP Publishing},
volume = {16},
number = {1},
pages = {013015},
author = {Orus, Roman and Wei, Tzu-Chieh and Buerschaper, Oliver and Nest, Maarten Van den},
title = {Geometric entanglement in topologically ordered states},
journal = {New Journal of Physics}
}

@misc{SM,
    note = {In the supplementary materials, we present a detailed proof of $\mathcal{H}_{\chi=2} \subseteq \mathcal{H}_{D=1}$, together with the full description for the computation of the negative logarithmic fidelity $F$ and the matrix-product entanglement $E_{\chi}$. We also provide the derivation of the upper bounds of virtual dimension ($\chi=2^{D}$) based on the exponential increase of the virtual dimension in the implementation of a $D$-layer quantum circuit to an MPS.}
}

@article{doi:10.1137/090752286,
author = {Oseledets, I. V.},
title = {Tensor-Train Decomposition},
journal = {SIAM Journal on Scientific Computing},
volume = {33},
number = {5},
pages = {2295-2317},
year = {2011},
doi = {10.1137/090752286},

URL = { 
    
        https://doi.org/10.1137/090752286
    
    

}
}

\end{document}